\documentclass[a4paper,11pt]{article}
\usepackage{jcappub} 
\usepackage{aas_macros}
\usepackage{amssymb}
\usepackage{amsmath}
\usepackage{graphicx}
\usepackage{dcolumn}
\usepackage[numbers]{natbib}
\usepackage{hyperref}
\usepackage{color}
\usepackage{ulem}
\usepackage{orcidlink}
\title{Entropy Evolution Towards DARKexp of IllustrisTNG Dark Matter Halos}

\author[a,1\orcidlink{0009-0006-5681-6297}]{A. Francis,\note{Corresponding author.}}
\author[a\orcidlink{0000-0002-6039-8706}]{L. L. R. Williams,}
\author[b\orcidlink{0000-0002-4571-2306}]{J. Hjorth}

\affiliation[a]{School of Physics and Astronomy, University of Minnesota, \\ 116 Church Street SE, Minneapolis, MN 55455, USA}
\affiliation[b]{DARK, Niels Bohr Institute, University of Copenhagen, \\ Jagtvej 155A, 2200 Copenhagen, Denmark}

\emailAdd{franc819@umn.edu}
\emailAdd{llrw@umn.edu}
\emailAdd{jens@nbi.ku.dk}

\abstract{Dark matter (DM) halos simulated via N-body techniques are known to exhibit nearly universal profiles at equilibrium; however, their origin remains uncertain despite thorough investigation. This work aims to probe the origin of simulated DM halo structure by testing DARKexp, a first-principles approach that describes the maximum entropy state of collisionless, self-gravitating systems in equilibrium via their energy distributions, and proposes an entropy functional that is expected to increase during evolution. We fit the DARKexp energy distribution to a set of massive equilibrium halos from the cosmological simulation IllustrisTNG. For the first time, we calculate the entropy of these halos as a function of cosmological time by tracking halos that are in equilibrium at $z = 0.0$, to $z = 3.0$ and calculating their entropy at various epochs between. We find that DARKexp provides an excellent fit to the energy distributions of equilibrium DM halos and that such halos exhibit an overall increase in entropy during evolution. Our results indicate that DM halos evolve to become their maximum entropy state at equilibrium and that this state is described by DARKexp.}

\keywords{dark matter simulations, dark matter theory}

\begin{document}
\maketitle
\flushbottom

\section{Introduction}
\label{sec:intro}
\indent \indent Since the discovery of its existence, dark matter (DM) has remained one of the most intriguing mysteries of the universe to particle physicists, astrophysicists and cosmologists alike. Some of the earliest works suggesting the existence of DM were published in the 1930s and discussed results from galaxy cluster observations \citep{Zwicky1993,Smith1936}. However, it was not until the late 1970s that this idea became popularized through Vera Rubin’s discovery of flat galactic rotation curves in the optical and observations of HI gas in the radio by Morton Roberts and Robert Whitehurst \citep{Rubin1978,Roberts1975}. Since then, new observations have provided strong evidence that DM not only exists, but makes up the majority of matter in the Universe: temperature fluctuations in the Cosmic Microwave Background (CMB), Big Bang Nucleosynthesis (BBN), gravitational lensing, etc. Cosmological simulations that model the evolution of DM and/or baryons over cosmological timescales have also provided evidence for DM by recreating the large-scale structure observed in the Universe \citep{Davis1985}. Today, with evidence from observations and simulations, it is widely accepted that DM constitutes roughly 85\% of matter and sets the building blocks for the large-scale structure observed in our Universe \citep{Gianfranco2018}.

Despite the fact that our universe is composed mostly of DM, it eludes detection and cannot be studied directly; thus, scientists turn to indirect methods to better understand the nature of DM. Cosmological simulations, such as N-body and hydrodynamical simulations, allow scientists to indirectly investigate the creation and evolution of DM structure on scales from individual halos to large cosmological volumes. N-body techniques simulate gravitational interactions and are often used to simulate dark-matter-only (DMO) mock Universes. Hydrodynamical techniques simulate baryonic processes and are utilized in studies of DM when investigating the interaction between baryons and DM. As it turns out, most of our knowledge about the equilibrium structure of DM halos has been derived from N-body simulations. The first simulations of cold dark matter (CDM) took place in the 1980s \citep{Davis1985,Dubinski1991}, and since then computational advancements have greatly improved simulations, allowing numerous studies to utilize them to better understand the nature of DM.

One of the most intriguing results gathered from simulations is that of the equilibrium DM halo (near) universal density profile, discovered in the 1990’s and fit by Navarro, Frenk and White (NFW) \citep{NFW1996,NFW1997}. The NFW fitting function is,
\begin{equation}
    \label{eqn:eqn1}
    \rho(r) = \frac{\rho_s}{(r/r_s)[1+(r/r_s)]^2}.
\end{equation}
The NFW profile is roughly isothermal over most radii, with a slope shallower than $-2$ toward the center of the halo and steeper than $-2$ near the virial radius. The characteristic radius, $r_s$ is the radius where the slope is exactly $-2$, and $\rho_s$ is the density at $r_s$. NFW is often adopted in many studies as the density profile for DM halos, however, the Einasto profile \citep{Einasto1965,Einasto1969}, a three-parameter fitting function, has been found to fit DM halos better than the two-parameter NFW profile \citep{Navarro2004}. The Einasto fitting function is,
\begin{equation}
    \label{eqn:eqn2}
    \rho(r) = \rho_se^{-2n[(\frac{r}{r_s})^{\frac{1}{n}}-1]}.
\end{equation}
The Einasto profile was first introduced in the 1960’s as a galactic stellar brightness volume density profile, but has proven to be useful for describing the density profiles of DM halos \citep{Navarro2004,Navarro2010}. Compared to the NFW profile, the Einasto profile adds an additional parameter, $n$, which is referred to as the Einasto index, and determines the slope steepness. Another common way to write this equation uses $\alpha$, which is just $\frac{1}{n}$. The larger the value of $\alpha$, the more rapidly the slope changes with radius.

Theoretical explanation of these near universal profiles and their shape has a long history. Scientists have tried to explain these profiles along with overall halo structure using various approaches. On the phenomenological side, \citep{Syer1998} argued that the profiles are a direct result of hierarchical merging while \citep{Huss1999} found that isolated halos also exhibit the known near universal profiles, suggesting that the profiles are not a direct consequence of hierarchical merging. \citep{Lu2006} suggested that a two-phase halo assembly is responsible for the universal profiles, reasoning that an initial fast accretion phase is responsible for the inner slope while a later, slower accretion phase results in the outer slope. Similarly, \citep{Ascasibar2007} found that the secondary infall model resulted in structures similar to those of simulated halos. 

Other studies have used a first-principles based approach rather than phenomenological to describe DM halo structure. Most studies utilizing this approach, including this work, argue that the equilibrium state of a halo corresponds to its maximum entropy state. The first work to use such an approach was \citep{Ogorodnikov1957}, who arrived at the distribution function that has the Maxwell-Boltzmann form, $f(E)\propto\exp(-\beta E)$, where $E$ is particles' energy. Ten years later, \citep{Lynden-Bell1967} attempted the maximum entropy approach again, this time incorporating the collisionless nature of stellar systems. In the limit appropriate for galaxies, he arrived at the same Maxwell-Boltzmann distribution, $f(E)$. Systems resulting from such $f(E)$ are isothermal spheres and do not resemble observed galaxies. Later authors attempted to reconcile this by assuming that real galaxies do not reach a state described by maximum entropy $f(E)$ because they are somehow prevented from it, often referred to as incomplete relaxation \citep{Stiavelli1987,Hjorth1991,Spergel1992}. Various models of incomplete relaxation have been proposed; for example, \citep{Pontzen2013} postulated that relaxation is incomplete because properties other than energy are also conserved within a collapsing system; specifically, the angular and radial actions. Their final distribution, $f(E,\bf{J})$, was different from the Maxwell-Boltzmann form and had a total of four fitted shape parameters. Although their results exhibited a significant improvement from \citep{Lynden-Bell1967}, they indicate that the central regions of halos would require a separate population of particles; thus, an important constraint is missing from their analysis.

Binney \citep{Binney1982} was the first to suggest that $f(E)g(E)=n(E) \sim \exp(-\beta E)$, resulting in galactic profiles that obeyed the observed $r^{\frac{1}{4}}$ mass-density profile. (Here, $g(E)$ is the density of states, and $E$ is the sum of kinetic and potential energy.) Keeping to the first principles based approach and furthering the ideas of \citep{Binney1982} to DM halos, this paper tests DARKexp, the energy distribution predicted via the statistical mechanical theory of collisionless self-gravitating systems with isotropic velocity dispersion, outlined in \citep{Hjorth2010}, and extended to include entropy evolution by \citep{Williams2022}.
The theory is similar to the standard derivation of the Boltzmann distribution with two changes: (1) instead of computing the number of possible microstates in phase space, it is computed in energy space, and (2) low occupation numbers are important and therefore treated more accurately than with Stirling’s approximation. The result of these two changes is the differential energy distribution for systems in equilibrium, DARKexp,
\begin{equation}
    \label{eqn:eqn3}
    n(E) = A(e^{-\beta[E - \Phi_0]} - 1) = A(e^{[\phi_0-\epsilon]} - 1),
\end{equation}
where $n(E)$ is the number of particles with energy $E$, $\beta$ is the inverse temperature, and $A$ is a number determined by the mass of the system. $E$ and $\Phi_0$ have units of energy. DARKexp deviates from the Maxwell-Boltzmann distribution and has been previously tested and found to fit well the energy distributions and density profiles of simulated equilibrium halos \citep{Williams2010,Hjorth2015,Nolting2016}, and density profiles of observed clusters \citep{Beraldo2013,Umetsu2016}. Putting all this in perspective, DARKexp is arguably the best statistical mechanical theory to describe DM halo structure.

Utilizing the derivation of DARKexp, \citep{Williams2022} postulated an entropy-like functional that, theoretically, should increase during halo evolution and eventually reach a maximum at equilibrium. The entropy functional is,
\begin{equation}
    \label{eqn:eqn4}
    S_D = -\!\!\int \ln \{ \Gamma[n(E)] \} dE - \beta\!\!\int n(E)EdE.
\end{equation}
$S_D$ yields DARKexp, \eqref{eqn:eqn3}, when extremized.

The central aim of this paper is to test whether $S_D$ increases during the evolution of simulated DM halos. Previously, $S_D$ was shown to increase monotonically for very simplified, spherically symmetric halos \citep{Williams2022}, however, the critical test can only be carried out with high-resolution N-body simulations. Here, we use the state-of-the-art cosmological simulation IllustrisTNG. Before testing $S_D$, we fit DARKexp, \eqref{eqn:eqn3}, to equilibrium halos at the present epoch to ensure DARKexp accurately describes their energy distributions.

This paper is outlined as follows. Section II describes the simulation used. Section III discusses the methods used to determine equilibrium halos. Section IV details the procedure used to calculate density profiles, velocity anisotropies, energy distributions, and DARKexp fits for our set of equilibrium halos and analyzes the results. Section V outlines the technique used to track halos from $z = 0.0$ to $z = 3.0$ and Section VI discusses the evolution of energy distributions and entropy across the tracked redshifts.

\section{Simulation}
\begin{figure*} [b!]
        \includegraphics[width=1\textwidth]{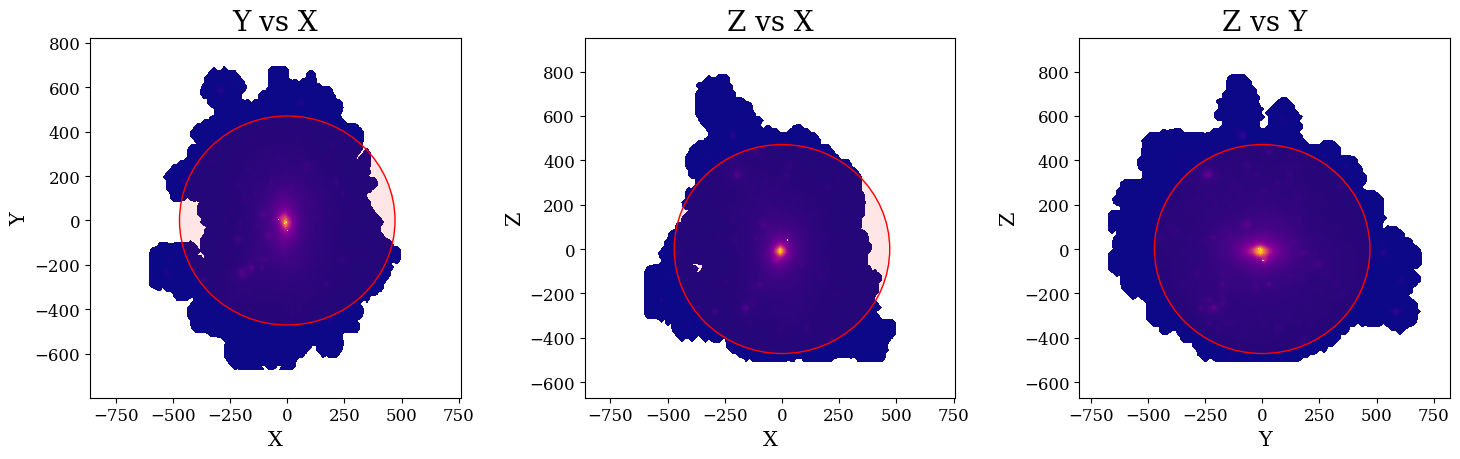}
        \includegraphics[width=1\textwidth]{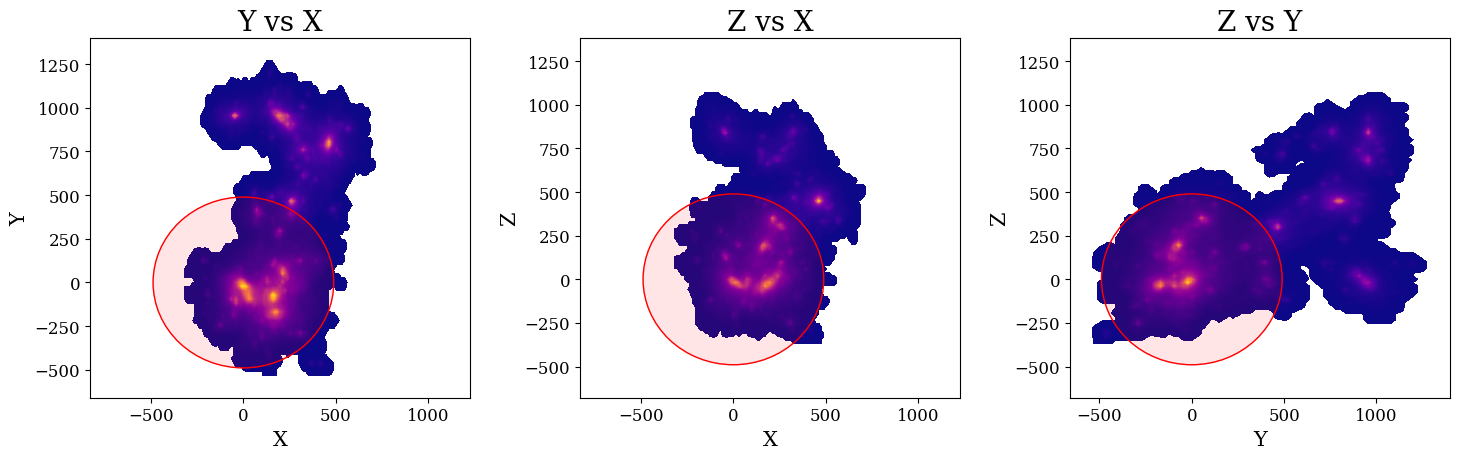}
        \caption{Heat maps of two systems defined as halos in TNG50-1-Dark. Colors correspond to linear density of particles, with purple and yellow revealing areas of low and high density respectively. The maps are calculated for X-Y, X-Z, and Y-Z grids individually. Grids were created in a given plane by representing the space around the halo as a grid of roughly $20\,$kpc$^2$ squares, each with a specific number of particles belonging to it. The least dense areas (outermost purple areas) correspond to roughly 1-10 particles/($20\,$kpc$^2$) while the most dense areas (innermost yellow areas) correspond to $~10^6$ particles/($20\,$kpc$^2$). The red shaded circle shows the virial radius centered at the point of deepest potential. {\it Top Panels:} Example of an equilibrium halo which corresponds to group number 19 at snapshot 99 ($z = 0.0$), or Halo F in our equilibrium halo set. {\it Bottom Panels:} Example of a halo considered not to be in equilibrium which corresponds to group number 7 at snapshot 50, or $z = 1.0$.}
        \label{fig:fig1}
\end{figure*}
\indent \indent The data analyzed in this paper comes from the DM-only counterpart to IllustrisTNG’s TNG50 simulation- TNG50-1-Dark \citep{Nelson2019}. The IllustrisTNG project is a suite of cosmological hydrodynamical simulations composed of three main simulations with different volumes. It is the successor, or The Next Generation (TNG), of the Illustris project \citep{Genel2014}. Like its predecessor, IllustrisTNG (hereafter, TNG) runs on the moving-mesh code AREPO but with an updated galaxy formation model \citep{Springel2010}. The updated galaxy formation model includes a new model for SMBH growth and AGN feedback that updates black-hole-driven feedback injection at low accretion rates. Cosmic magnetism is also newly treated and various numerical improvements were made in TNG \citep{Pillepich2018}. Because we utilize the dark-matter only counterpart to the full TNG50 simulation, the updated galaxy formation model and newly-treated cosmic magnetism incorporated into TNG are not used here but noted to illustrate the level of detail in the TNG simulations.

Along with the new physics and numerical improvements, TNG has surpassed Illustris by running three different cubic volume simulations- TNG50, TNG100 and TNG300, each with physical cubic volumes of roughly 50, 100, and 300 Mpc$^3$ respectively. Each cubic volume is run at varying resolution levels, the highest resolutions, denoted as TNG50-1, TNG100-1, and TNG300-1, are 2 x 2160$^3$, 2 x 1820$^3$ and 2 x 2500$^2$ respectively. A DM-only (DMO) counterpart is run at the highest resolution for each simulation. Because TNG50-1 has the highest resolution and smallest volume, it offers the best resolution of individual halo structure.

TNG50-1-Dark, along with the other simulations, is run starting at a redshift of $z=127$ to present day $z=0.0$ with Newtonian gravity solved in an expanding Universe. The initial conditions are motivated by a cosmology consistent with Planck Collaboration (2016) results ($\Omega_{\Lambda,0}$ = 0.6911, $\Omega_{m,0}$ = 0.3089, $\Omega_{b,0}$ = 0.0486, $\sigma_8$ = 0.8159, $n_s$ = 0.9667 and $h$ = 0.6774). The simulation includes 2160$^3$ DM particles each with mass 3.647556e5 \(M_\odot\)/$h$. Structures are identified using a standard friends-of-friends (FoF) algorithm with linking length $b$ = 0.2 and Plummer equivalent gravitational softening at $z = 0.0$ of 0.29 kpc. There are 100 snapshots stored for each simulation, each corresponding a specific redshift of the simulation. Snapshot information holds details about individual particles and each snapshot file has a corresponding group catalog with details about halos and subhalos. Group catalogs hold data such as halo/subhalo mass, radius, center of mass, etc. Out of the 100 snapshots, only 20 are considered full and the rest are considered mini snapshots. In this work we gather data from full snapshots only.

While all halos identified are gravitationally bound systems, not all are isolated and many are merging systems with dense substructure found far from the particle of deepest potential. Figure~\eqref{fig:fig1} shows two systems identified as halos, the top image shows an isolated halo determined to be in equilibrium while the bottom shows a halo undergoing a possible merging event with multiple substructures. The requirements for equilibrium are discussed in the next section.

\section{Determining Equilibrium}\label{sec:equilibrium}

\indent \indent DARKexp, eq.~\eqref{eqn:eqn3}, is the energy distribution of fully relaxed, collisionless, self-gravitating and isotropic systems, therefore, it is tested only on halos in equilibrium. We determine equilibrium at present day in the simulation, $z = 0.0$. Furthermore, to guarantee that the halos are sufficiently resolved, we only analyze those with a mass greater than 10$^{13}$ \(M_\odot\)/$h$. To determine equilibrium we utilized two criteria published by \cite{Neto2007} that a halo must meet: $S$ value and virial ratio, $VR$. The $S$ value determines center of mass offset from deepest potential, 
\begin{equation}
    \label{eqn:eqn5}
    S = \frac{|r_{pos} - r_{CoM}|}{r_{200}},
\end{equation}
where $r_{pos}$ denotes the coordinates of deepest potential, $r_{CoM}$ denotes the coordinates of the center of mass and $r_{200}$ is the virial radius. In this calculation, and all subsequent calculations, $r_{200}$ is defined as $R_{TopHat200}$ in the group catalog data.  We use the value given in \citep{Neto2007} as our cutoff:  $S < 0.07$ corresponds to a halo in equilibrium. A higher $S$ value corresponds to a larger offset between deepest potential and center of mass, corresponding to a halo that has not yet reached equilibrium. Once halos with $S>0.07$ were thrown out, the $VR$ of the remaining halos was calculated using
\begin{equation}
    \label{eqn:eqn6}
    VR = -2 \frac{KE_{tot}}{PE_{tot}}.
\end{equation}
$PE_{tot}$ and $KE_{tot}$ denote the sum of particle potential and kinetic energies within $r_{200}$. The particle potential is given in the snapshot data while the kinetic energy required some calculation, which is described in Section V. The virial theorem states that equilibrium systems have a kinetic energy equal to half the negated potential energy, $<K> = -\frac{1}{2}<U>$. Therefore, systems with a $VR$ close to 1 are considered in equilibrium. Again, we use the values given in \cite{Neto2007} as our cutoff; halos are considered in equilibrium with a $VR$ between [0.9, 1.35]. A third criterion, subhalo mass fraction, was discussed in \cite{Neto2007}, however, we do not use this value in our criteria because TNG50-1-Dark halos were found to provide unreasonable numbers for this value. Rather than completely ignore substructure mass fraction, we performed a visual inspection of individual halos as 2d heat maps in the $x-y$, $x-z$, and $y-z$ axes (as in Figure~\eqref{fig:fig1}) to ensure no massive substructure was present in halos determined to be in equilibrium.

The final set consisted of 6 halos named A-F, A being the most massive and F the least massive. Table 1 shows the properties of these halos at $z = 0.0$, along with their calculated $S$ and $VR$ values.
\begin{table}[t!]
\centering
        \begin{tabular}{|c|c|c|c|c|c|c|}
         \hline
         \textbf{Halo}& \textbf{GrNr}&\textbf{Mass}&$\mathbf{r_{200}}$&$\mathbf{VR}$&$\mathbf{S}$ \\
         \hline
         A& 2 &5.19& 758.6 & 1.14 & 0.048 \\
         \hline
         B &  6 &3.00&632.1 &0.96&  0.056  \\
         \hline
         $\quad$C$\quad$&$\quad$ 7 $\quad$ &$\quad$2.80$\quad$&$\quad$617.4$\quad$ &$\quad$1.03$\quad$&  0.048  \\
         \hline
         $\quad$D$\quad$&$\quad$14$\quad$&$\quad$1.61$\quad$&$\quad$513.7$\quad$ &$\quad$0.96$\quad$&  0.039  \\
         \hline
         $\quad$E$\quad$&$\quad$16$\quad$&$\quad$1.35$\quad$&$\quad$483.9$\quad$&$\quad$0.90$\quad$&  0.048  \\
         \hline
         $\quad$F$\quad$&$\quad$19$\quad$&$\quad$1.24$\quad$&$\quad$471.0$\quad$&$\quad$1.06$\quad$&  0.048  \\
         \hline
    \end{tabular}
    \caption{Individual Halo Properties at $z=0.0$ for our set of equilibrium halos. We present our naming scheme of equilibrium halos (A-F) in the first column and the group number (GrNr) associated with each halo at $z = 0.0$ in second column. Mass is the total mass enclosed within $r_{200}$ in units of ($10^{13}$\(M_\odot\)). $r_{200}$ is the virial radius in units of$kpc$. The last two columns are the values obtained via our equilibrium determination, $VR$ stands for virial ratio and $S$ is the value that determines offset from deepest potential.}
\end{table}
\begin{figure*}[h!tbp]
    \centering
    \includegraphics[width=1\textwidth]{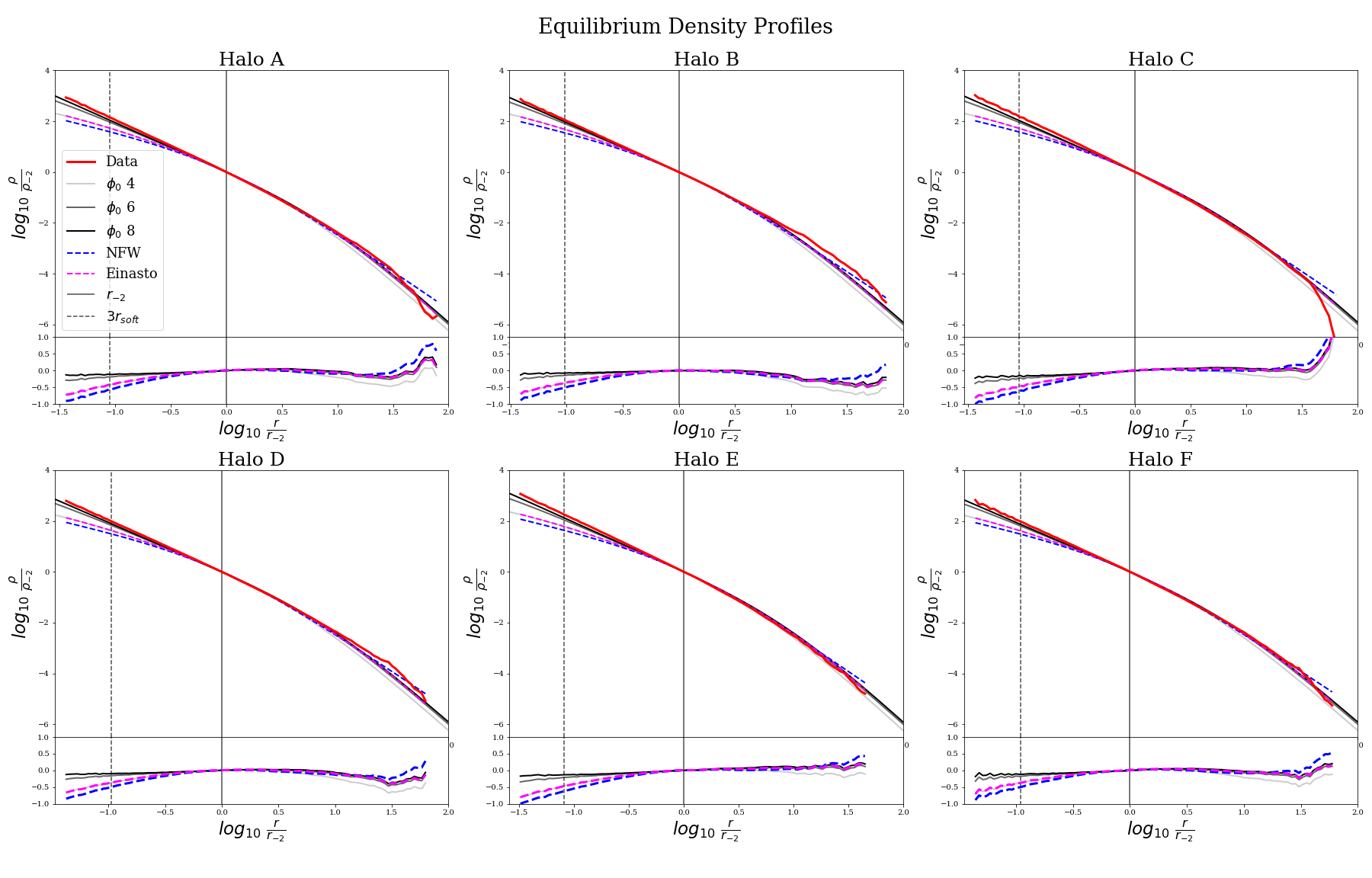}
    
    \caption{Density profiles of equilibrium halos at $z=0.0$ and their NFW, Einasto, and DARKexp fits. Residuals for the fits are shown below each density profile. The x-axis of each plot is the log of the radius in units of $r_{-2}$, the radius where the slope = $-2$. The y-axis is the log of the density in units of $\rho(r_{-2})$. The red line shows the data and black shows DARKexp fits at three values of $\phi_0$ -- $4.0$, $6.0$ and $8.0$ -- with a darker black corresponding to a higher value of $\phi_0$. The blue and magenta dashed lines show the NFW and Einasto fits respectively. Einasto was fit using $\alpha$ = 0.17 and both fits use the same scale radius, $r_{-2}$. A radius corresponding to $3r_{\rm soft}$ ($3\times 0.29$kpc) is shown with a grey dashed line and the grey solid line shows 0.0 on the x-axis ($r_{-2}$) for reference.}
    \label{fig:fig3}
\end{figure*}

\begin{figure*}[h!tbp]
    \centering
    \includegraphics[width=1\textwidth]{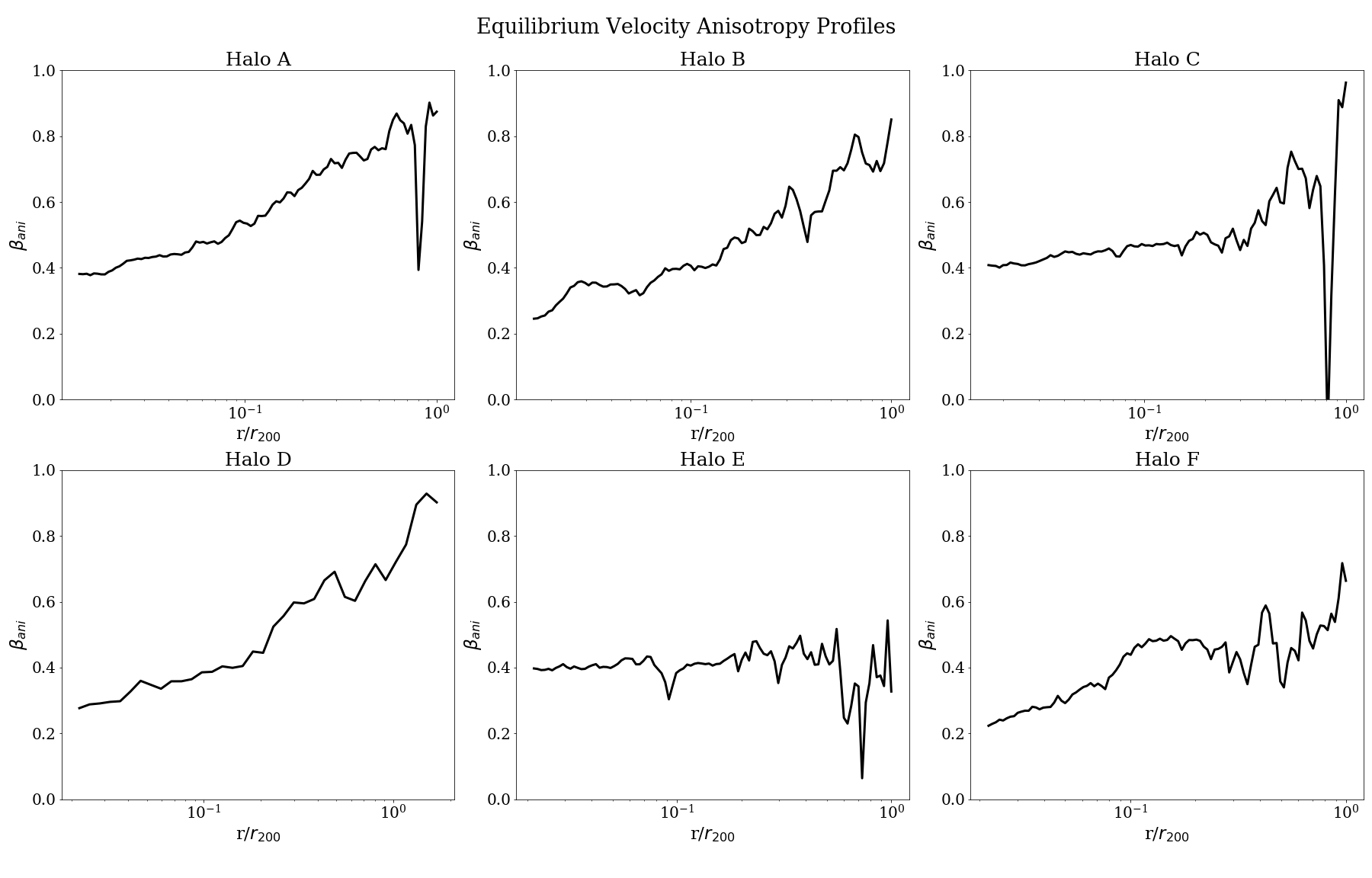}
    
    \caption{Velocity anisotropy profiles of our set of equilibrium halos at $z=0.0$. The y-axis is the velocity anisotropy, $\beta_{ani}$, and the x-axis is the log of the radius in units of $r_{200}$.}
    \label{fig:fig4}
\end{figure*}

\section{Analyzing Equilibrium Halos}

\indent \indent \textbf{Density Profiles.~~}
The density profile of each equilibrium halo was calculated at $z = 0.0$. To do this, the particles were binned in log-spaced spherical shells centered on the particle of deepest potential. The radial range was defined the same for each halo, with the inner radii set to 0.3 kpc and the outer radii to $r_{200}$. We chose the inner radii to be just greater than the softening radius, which is $r_{\rm soft} = 0.29$ kpc at $z = 0.0$. The density profiles for our set of equilibrium halos can be found in Figure \eqref{fig:fig3}.

The density profiles were fit with the NFW and Einasto fitting functions, Equations \eqref{eqn:eqn1}, and \eqref{eqn:eqn2}. The best-fit value of the scale radii, $r_s$, for the Einasto profiles were extremely small values, on the order of $10^{-3}$ $r_{200}$, but manually varying $r_s$ had little effect on the density profile fits. For simplicity, we adopted a value of $r_s$ = $r_{-2}$ for all profiles, where $r_{-2}$ is the radius where the density profile has a slope of $-2$. The value of the Einasto index used for the Einasto fit, $n$ in Equation \eqref{eqn:eqn2}, was motivated by \citep{Navarro2004} who found an average value of $n = 5.88$ ($\alpha = 0.17$) for halos of various masses.

The DARKexp density distribution does not have an analytic form; therefore, we used pre-calculated DARKexp density profiles with $\phi_0 = 0.5$ to $8.0$ in $\Delta\phi_0 = 0.1$ increments. The density profiles calculated from the TNG50 simulation were scaled to match the scaling of the DARKexp profiles: such that ($\log_{10}\frac{r}{r_{-2}}$,$\log_{10}\frac{\rho(r)}{\rho_{-2}}$) corresponds to $(0.0, 0.0)$ on the density plot. The same scaling was done for the NFW and Einasto fits.

The NFW, Einasto and DARKexp profiles fit well to all halos in our set, which is expected of halos in equilibrium. The Einasto profile fits our data better than the NFW at both small and large radii, however, neither fit our data at innermost radii. This steep inner slope is found for all halos in our set and has been found in other papers utilizing IllustrisTNG TNG50-DMO simulations, for example, Figure 1 in \citep{Gross2024}. DARKexp fits this steep inner slope better than NFW and Einasto with larger values of $\phi_0$. The best-fitting DARKexp profile was calculated by finding the root-mean-square deviation and corresponded to the $\phi_0$ = $8.0$ for each halo. $\phi_0$ values greater than $8.0$ are almost indistinguishable from $\phi_0$ = $8.0$. We do not comment on the origin of this inner slope because the purpose of these profiles in this work is not to explain small features such as this but rather to confirm our selection of equilibrium halos. 

\textbf{Bulk Velocity.~~}
Each halo is subject to its own movement through the mock Universe. To calculate the bulk velocity, the velocities of the innermost particles, those within $0.5r_{200}$, were averaged. We then subtracted the bulk velocity from individual particle velocities before using those values in our calculations of kinetic energy and velocity dispersion.

\textbf{Velocity Anisotropy.~~}
The velocity anisotropy was calculated for each halo at $z=0.0$. The particles were binned in log-spaced spherical shells with inner radii set to 1 kpc and outer radii to $r_{200}$. The average tangential and radial velocities of each shell were calculated using
\begin{equation}
    \label{eqn:eqn7}
    \bar{v}_{r} = \frac{\sum_{i=1}^{N} \vec{v_i} \cdot \hat{r_i}}{N},
\end{equation}
\begin{equation}
    \label{eqn:eqn8}
    \bar{v}_{t} = \frac{\sum_{i=1}^{N} | \vec{\omega_i} \times \vec{r_i} |}{N},
\end{equation} 
where subscript $i$ refers to individual particles. Their total number in each spherical shell, $N$, is different for different shells. The radial and tangential velocity dispersions of each shell were then calculated using
\begin{equation}
    \label{eqn:eqn9}
    \sigma_{r} = \sqrt{\frac{\sum_{i=1}^{N} [\vec v_i\cdot \hat r_i - \bar{v}_{r}]^2}{N}},
\end{equation}
\begin{equation}
    \label{eqn:eqn10}
    \sigma_{t} = \sqrt{\frac{\sum_{i=1}^{N} [| \vec{\omega_i} \times \vec{r_i} | - \bar{v}_{t}]^2}{N}}.
\end{equation} 
And finally, the velocity anisotropy of each shell was calculated using
\begin{equation}
    \label{eqn:eqn11}
    \beta_{ani} = 1 - \frac{\sigma_t^2}{2\sigma_r^2}
\end{equation}
$\beta_{ani}$ is velocity anisotropy here, not the fitting parameter for DARKexp, $\beta$.

\begin{figure*}[h!tbp]
    \centering
    \includegraphics[width=0.8\textwidth]{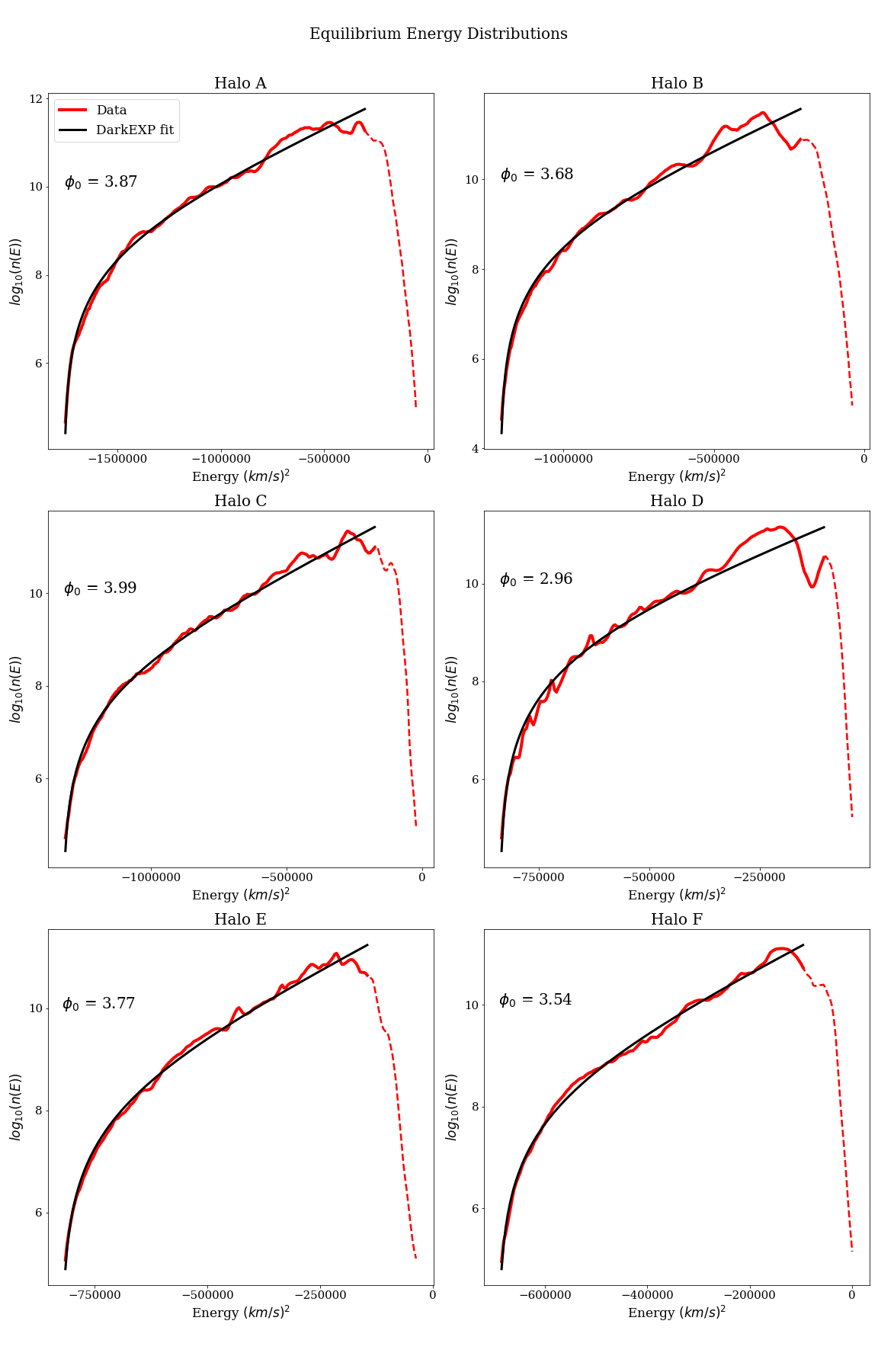}
    
    \caption{The energy distributions of our set of equilibrium DM halos at $z=0.0$. 
    The plot shows the particle energies on x-axis and log of number of particles on y-axis. 
    The solid red line shows data that was truncated just beyond the peak and used for fitting 
    DARKexp. The dashed red line shows the data from all particles belonging to the halo within it's virial radius. The black line is DARKexp's best fit, with the dimensionless $\phi_0$ annotated on each plot.}
    \label{fig:fig5}
\end{figure*}

The velocity anisotropy profiles of our set of halos, Figure~\eqref{fig:fig4}, resemble those found in other works such that the profiles are generally more tangential toward inner radii (smaller $\beta_{ani}$) and become more radial at outer radii (larger $\beta_{ani}$) \citep{Rasia2004}. The dips in $\beta_{ani}$ seen in Halos A and C at large radii are most likely due to substructure. Halo E is unique in that its $\beta_{ani}$ remains roughly constant across all radii, which we attribute to an unknown feature of its morphology. All of our halos are more radially anisotropic toward the center than expected. Although the reason for this is beyond the scope of this paper, one possibility is that in-falling particles that have not yet reached a steady orbit could be contributing to the velocity dispersion, causing our velocity anisotropy profiles to appear more radially anisotropic than expected. Overall, we see a general increase in $\beta_{ani}(r)$ from inner radii to outer radii for halos A, B, C, D and F, providing additional evidence that these halos are in equilibrium.

\textbf{Energy Distributions.~~}
The energy distributions were calculated as a histogram of individual particle energies, which are a sum of potential and kinetic energies. Setting $\Phi_0$ equal to the minimum potential energy of each halo at $z=0.0$, energy distributions were fit with DARKexp, Equation \eqref{eqn:eqn3}, by adjusting the fitting parameter $\beta$ and normalization constant $A$. The fits were done using {\tt scipy.optimize}'s {\tt curve\_fit} function \citep{2020SciPy-NMeth}. The energy distributions for our set of equilibrium halos and the corresponding fits can be found in Figure \eqref{fig:fig5}.

All our energy distributions resemble those found in previous works dealing with highly-resolved N-body simulated DM halos, further supporting our claim that these halos are in equilibrium \citep{Gross2024, Londrillo1991}. The energy distributions are slightly noisy, which is to be expected due to the substructure that can be resolved with N-body simulations. DARKexp fits very well, except near the escape velocity and further because DARKexp does not predict distribution of unbound particles. We truncate the data used to fit DARKexp at large radii to ensure that the fit does not include unbound particles in the fit. The dimensionless $\phi_0$ values fit for each halo are small, single-digit numbers, consistent with the $\phi_0$ values found for halos in Millennium Simulations and others \citep{Nolting2016,Hjorth2015}. 

\section{Tracking Halos from z=0.0 to z=3.0}
\indent \indent Due to the computational cost required to download merger trees, we track the equilibrium halos at previous redshifts (up to $z = 3.0$) using a simplified scheme outlined here. At each snapshot of the simulation, halos identified are assigned a group number ordered by mass, 0 being the most massive at that redshift. Because group number assignment is done each snapshot, individual halos can be assigned different group numbers at different redshifts. Individual particles, however, keep the same particle ID throughout the entire simulation. These group and particle assignments are done by the TNG team.
\begin{figure}[tb!]
    \centering
    \includegraphics[width=0.75\linewidth]{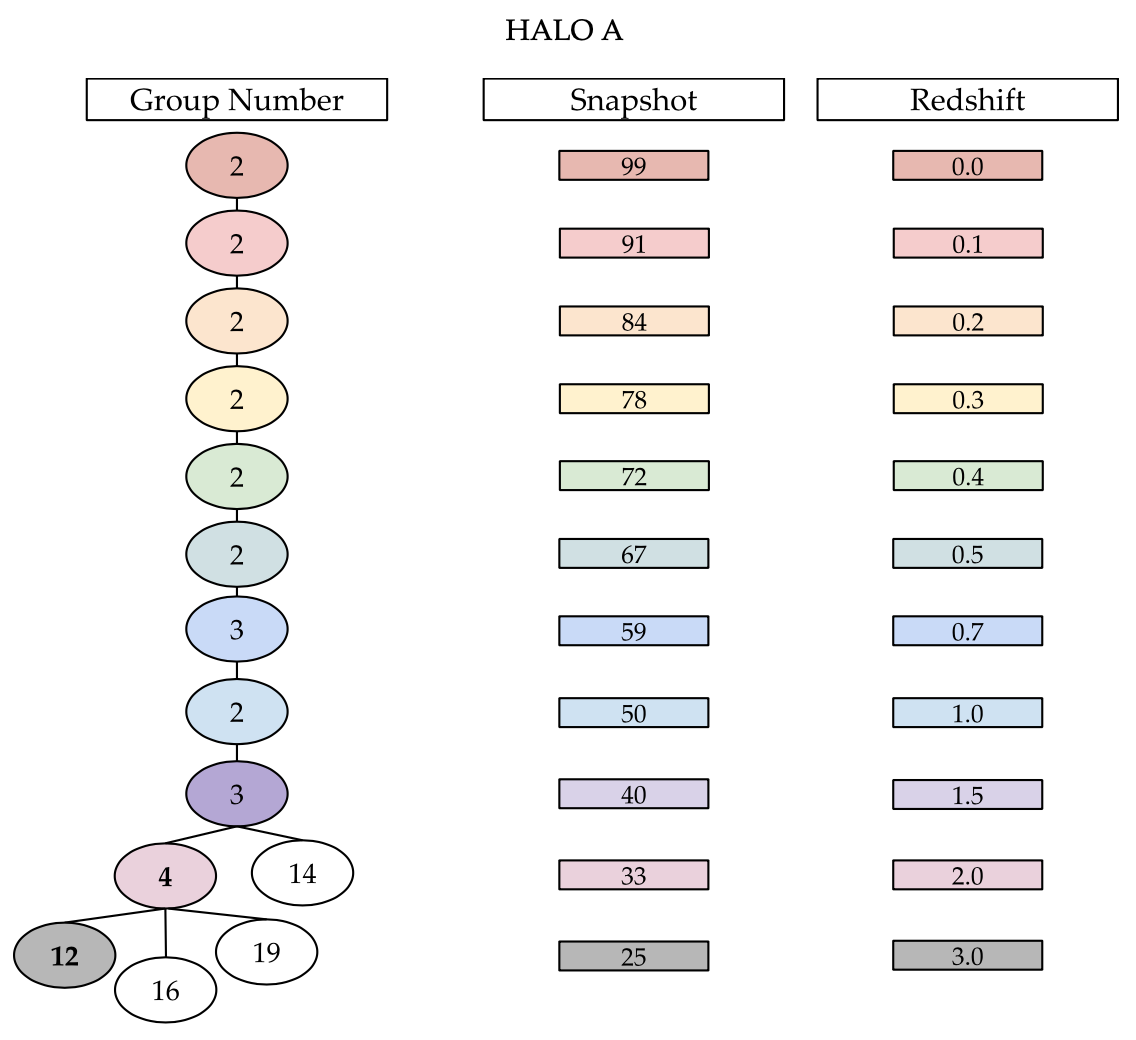}
    
    \caption{A schematic layout of the tracking process for Halo A. The snapshot number and corresponding redshift are color coded for each epoch. The group number that is assigned to Halo A at various epochs is found inside the ovals. The colored ovals represent the most massive ancestor halo at that epoch.}
    \label{fig:fig2}
\end{figure}

To track our set of halos back to $z = 3.0$, we utilized the full snapshots between $z = 0.0$ and $z = 3.0$. For a single halo in our set, we gathered the particle ID’s of all particles within the virial radius at $z = 0.0$. We then compared these particle ID’s with the particle ID’s of the first 100 halos in the previous snapshot ($z = 0.1$). If a halo in the previous snapshot contained over 50$\%$ of the same particles, it was determined to be the ancestor. This was repeated for each full snapshot going back to $z = 3.0$. Note that particle ID’s were only compared for full snapshots that were consecutive, that is, we compared $z = 0.1$ with  $z = 0.2$, and $z = 0.2$ with $z = 0.3$ and so on. This made certain we were tracking the evolution of a specific halo and not just the particles present in that halo at $z = 0.0$. 

Halos identified as separate structures at one epoch can merge and be identified as one halo at the next. When this occurs, we choose the ancestor to be the most massive halo identified in the previous full snapshot. A visualization of this technique can be found in Figure~\eqref{fig:fig2} where we track Halo A from  $z = 0.0$ to  $z = 3.0$. Merging events occurred between snapshot 33 and 40 and between snapshot 25 and 33 for this halo. We chose the most massive ancestor halo to be considered as the sole ancestor of Halo A at that epoch. We note that Halo E consisted of many extremely low-mass halos at $z=3.0$, as a result it was only able to be tracked back to $z=2.0$.

\section{Evolution of Energy Distributions and Entropy}

\begin{figure*}[!htp]
    \centering
    \includegraphics[width=0.85\textwidth]{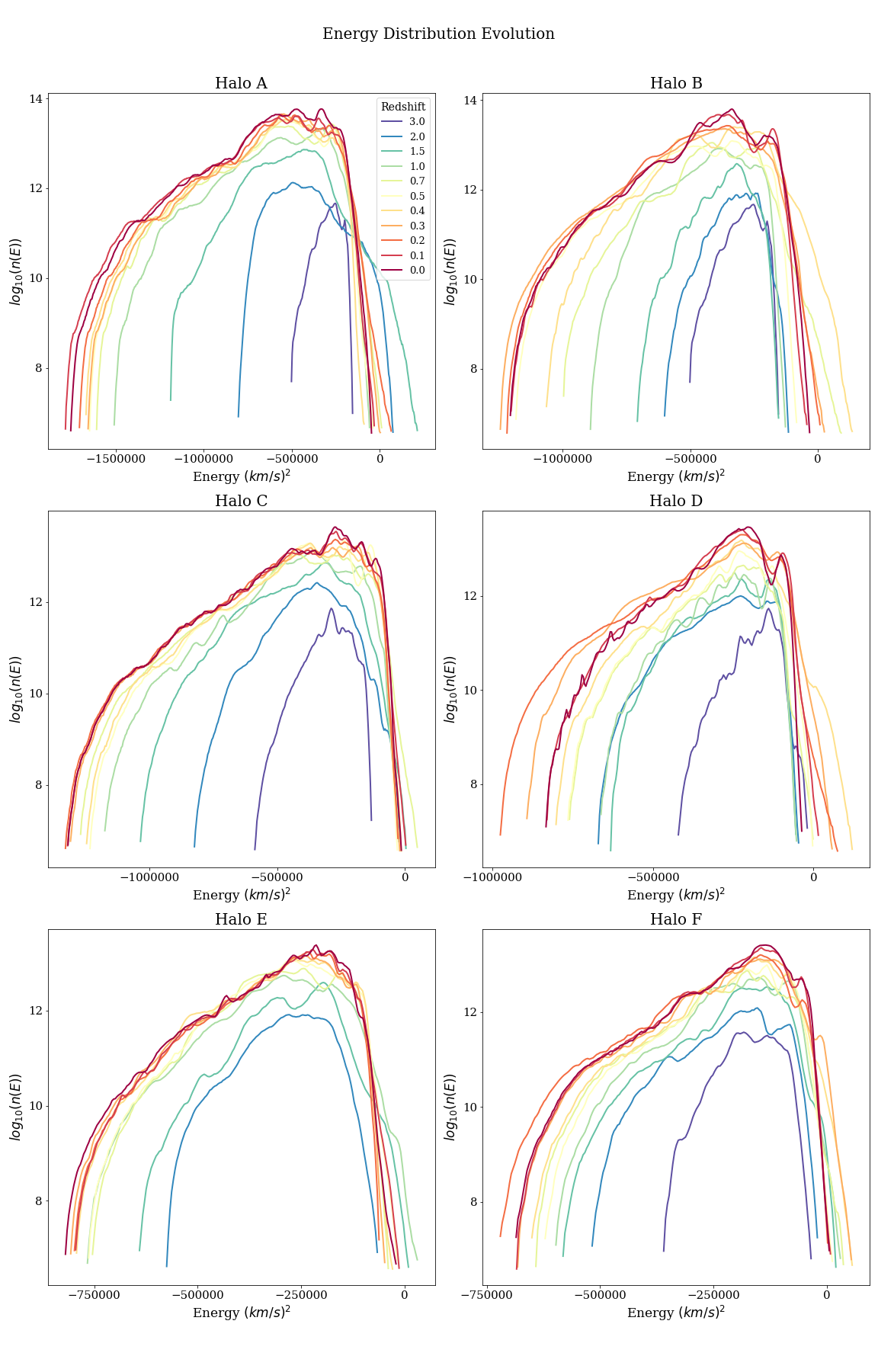}
    \caption{The energy distributions, $n(E)$, for our set of halos at multiple epochs. Each color denotes the redshift
    at that epoch, which can be found in the key in the upper left plot.}
    \label{fig:fig6}
\end{figure*}

\indent \indent This section details the crux of this work: investigating the energy distribution and entropy evolution of our set of equilibrium halos. Once equilibrium halos were tracked from $z=0.0$ to $z=3.0$, their energy distributions were calculated at each full snapshot between $z = 0.0$ and $3.0$ using the same technique described in Section IV. The energy distribution evolution for our set of halos can be found in Figure~\eqref{fig:fig6}. Note that because Halo E could not be tracked at $z=3.0$ it does not have an energy distribution for that epoch.

From Figure~\eqref{fig:fig6}, it can be seen that at early epochs all halos exhibit different energy distribution shapes compared to those at $z=0.0$. Moreover, the distributions at $z=3.0$ look quite different from halo to halo, suggesting that halos not in equilibrium can have various forms of energy distributions. Although different at early epochs, the energy distributions at equilibrium ($z=0.0$) all resemble DARKexp. The evolution of all halos' energy distribution towards DARKexp as they reach equilibrium indicates that DM halos evolve to become their maximum entropy state. 

We compare our results of the energy distributions, $n(E)$, at all epochs to those of halos generated via the Extended Secondary Infall Model (ESIM)---a spherically symmetric and isolated calculation of halo evolution \citep{Williams2022}. We see that they do not resemble each other at early epochs: those generated via ESIM have a large fraction of particles with significantly 
\clearpage
\begin{figure*}
    \includegraphics[width=1\textwidth]{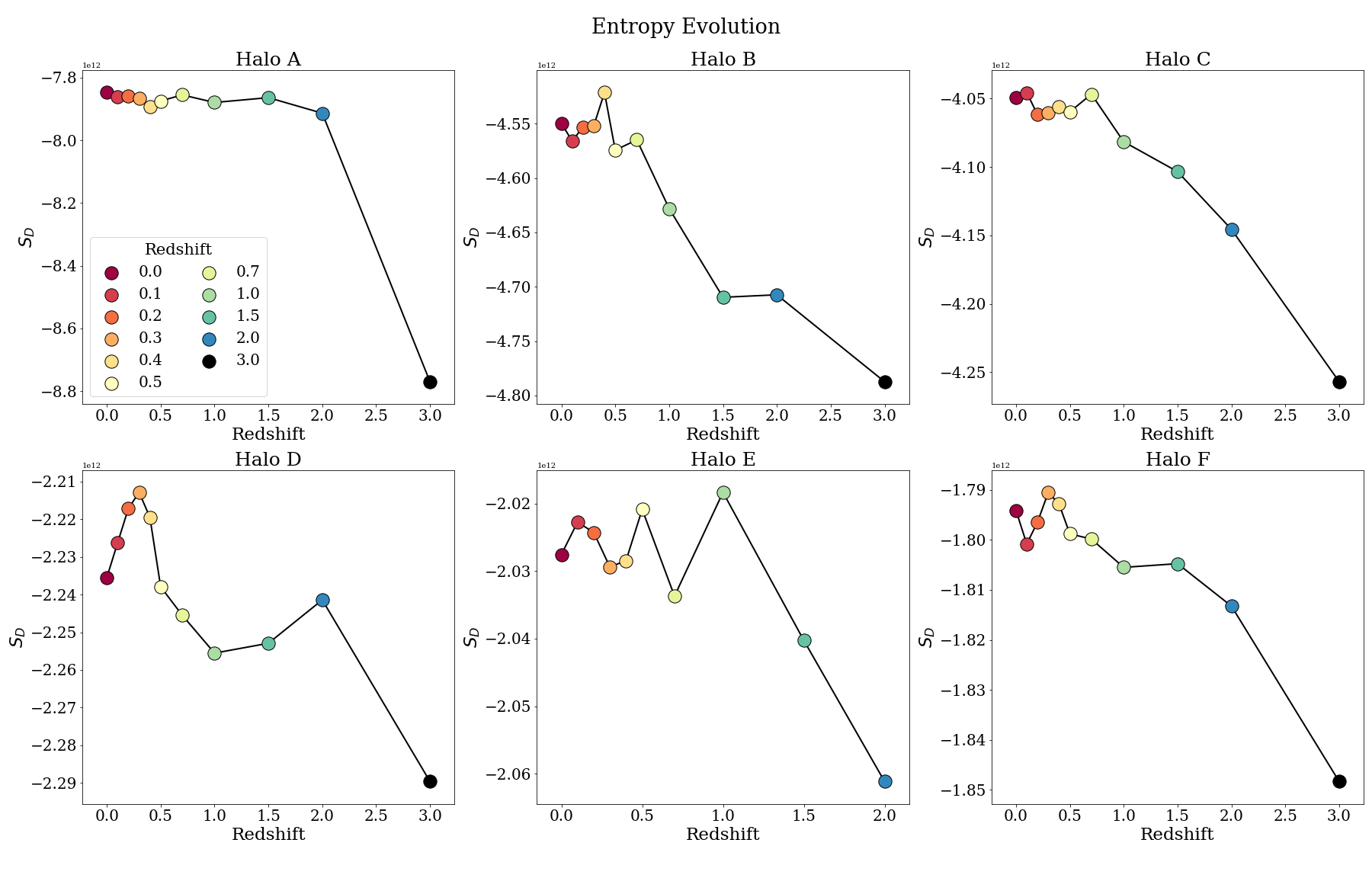}
    \caption{Evolution of halo entropy, $S_D$. Each colored circle represents a specific redshift at which the entropy is calculated, the key for the colors is shown in the upper left plot. The black solid line simply connects all the points.}
    \label{fig:fig7}
\end{figure*}
\noindent bound total energies, causing the energy distribution to peak towards more bound particles than what we see for the early epochs of halos generated in TNG. However, it is clear that $n(E)$ generated either way all broaden with time and evolve to become DARKexp. The differences at epochs prior to $z=0.0$ are most likely due to differences in the dynamical treatment of evolution in ESIM versus N-body.

Along with the energy distributions, the entropy of each halo was determined at each full snapshot between $z=0.0$ and $z=3.0$. To calculate entropy, we use Equation \eqref{eqn:eqn4}, which utilizes the energy distribution, $n(E)$, and $\beta$. $\beta$ is only determined at equilibrium, so all epochs previous to $z=0.0$ use the value of $\beta$ calculated for that halo at equilibirum. For a given halo, we normalized the energy distributions at epochs with $z>0.0$ to the energy distribution at $z=0.0$.

\begin{figure*}
    \includegraphics[width=1\textwidth]{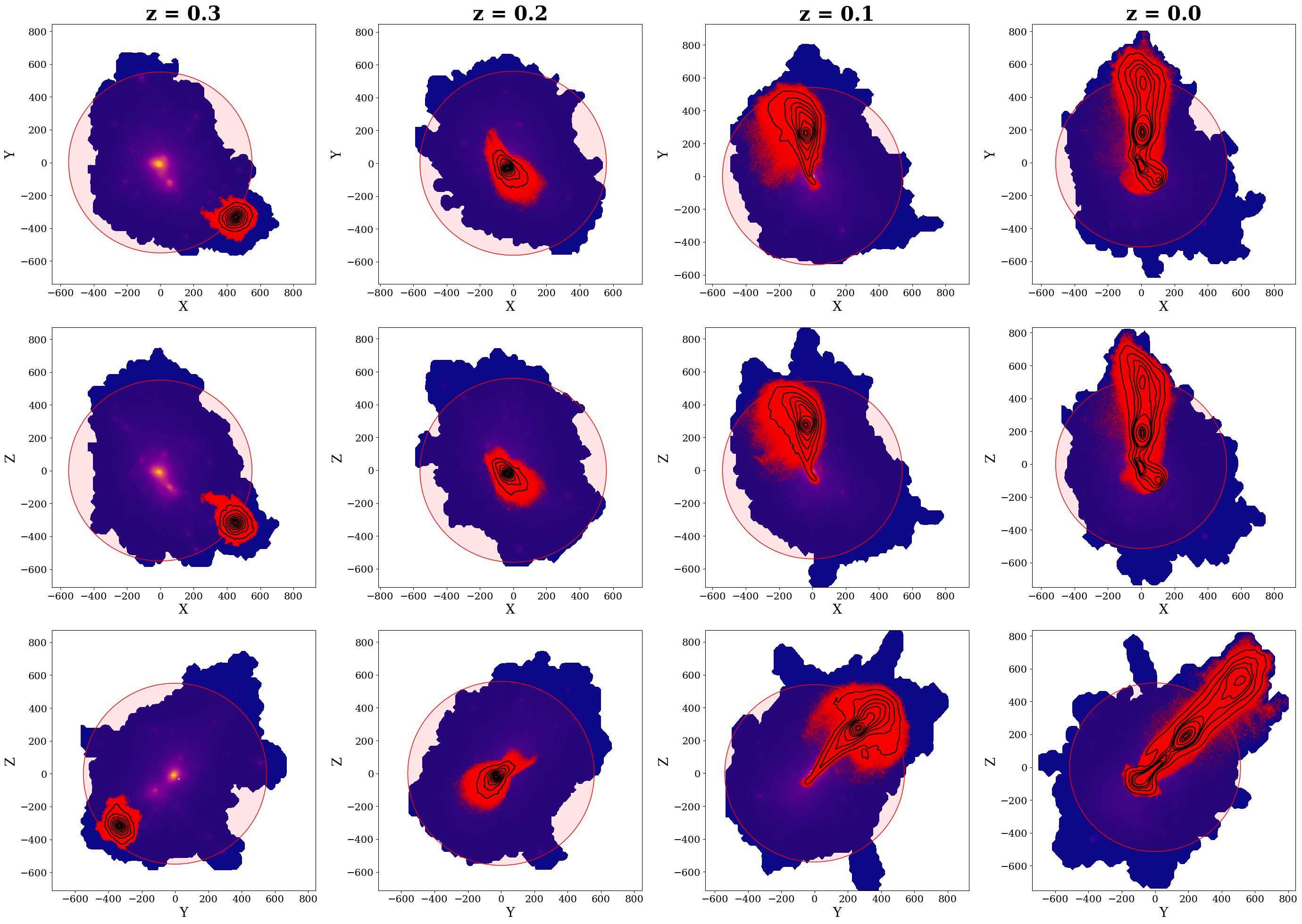}
    \caption{Heat maps of the projected particle density of Halo D at redshifts $0.3$, $0.2$, $0.1$, and $0.0$ from left to right. The three rows show the X vs. Y (top), X vs. Z (middle) and Y vs. Z (bottom) heat maps at each redshift. The heat maps were made in the same fashion as Figure~\eqref{fig:fig1} but over plotted by the particles belonging to the substructure (red) referenced in text along with black contour lines representing density. The substructure was found at $z=0.3$ by eye from the underlying heat map and identified as a subhalo belonging to Halo D. The IDs of particles belonging to the subhalo were used to plot their positions at each redshift.}
    \label{fig:fig9}
\end{figure*}

Our results for entropy evolution are shown in Figure~\eqref{fig:fig7}. These reveal that the entropy functional $S_D$, Equation~\eqref{eqn:eqn4}, generally increases as the halo evolves. We note that of the two terms in Equation~\eqref{eqn:eqn4} the first term dominates. The entropy does not increase monotonically, as it did in the idealized spherically symmetric ESIM halos \citep{Williams2022}, due to various merging events and kinematic influences that occur in N-body simulations that are not accounted for in ESIM. An example of this can be found in the entropy evolution for Halo D, particularly, there is a peak at $z = 0.3$ and subsequent decrease in entropies at $z= 0.2, 0.1,$ and $0.0$. We investigated this by examining heat maps of particle densities for Halo D at these epochs, shown in Figure~\eqref{fig:fig9}, and found that a small DM structure with large kinetic energy collides with the halo center between $z=0.3$ and $z=0.1$. The addition of this substructure to the center of Halo D resulted in the deepening of the point of deepest potential at $z=0.2$ and then as the substructure moved through the halo at $z=0.1$ and $z=0.0$ the point of deepest potential shifted toward less bound energies, therefore resulting in narrower $n(E)$ and causing the entropy to decrease. 

We also note that the entropy evolution looks somewhat different here than the ones presented in \citep{Williams2022} because of the total time lapse of the simulations. They calculate entropy back to $0.01$ of Hubble time whereas we calculate back to $z=3.0$, about $0.16$ of Hubble time. The values of entropy on the vertical axis of Figure~\eqref{fig:fig7} have a small range, roughly consistent with the change in $S_D$ for the ESIM halos at $0.16$ of Hubble time. 

To summarize, our results show that N-body simulated halos evolve to their equilibrium state, DARKexp, by maximizing the entropy functional $S_D$. 

\section{Conclusion}

\indent \indent Using N-body simulated DM halos, we have tested two main predictions of the first principles, maximum entropy based approach to collisionless self-gravitating systems, DARKexp: 1, the shape of the energy distribution at equilibrium, and 2, for the first time, the DARKexp entropy functional, $S_D$, that should increase as simulated halos evolve from $z = 3.0$ to equilibrium at the present epoch. 

We tested DARKexp on a set of six highly-resolved, massive DM halos from IllustrisTNG’s TNG50-1-Dark simulation. These six halos were determined to be in equilibrium at $z=0.0$ because they passed both the center offset $S$ Equation~\eqref{eqn:eqn5}, and virial ratio $VR$ Equation~\eqref{eqn:eqn6} equilibrium criteria. We analyzed their density profiles at $z=0.0$, Figure~\eqref{fig:fig3}, and found them to be fit well by both Einasto and NFW fitting functions, except at innermost radii where the density is greater than predicted by either fitting function. DARKexp was found to fit better than NFW and Einasto for $\phi_0 = 8$. The velocity anisotropies were also calculated for each halo at $z=0.0$, Figure~\eqref{fig:fig3}, and were found to generally increase as a function of radius. Despite the irregularities described in Section VI, the density and velocity anisotropy profiles of our set of halos provided further evidence they are indeed in equilibrium at $z=0.0$ and behave as expected. 

The main purpose of this work was to test DARKexp’s prediction for energy distribution and entropy functional. DARKexp’s best fit to the energy distribution of equilibrium halos at $z=0.0$, Figure~\eqref{fig:fig5}, is found to represent each halo very well. The energy distribution of each halo, which we tracked from $z=3.0$ to $z=0.0$, is found to evolve towards DARKexp, Figure~\eqref{fig:fig6}. Most notably, the entropy functional, $S_D$, derived from DARKexp has been shown to increase as a function of redshift, Figure~\eqref{fig:fig7}. These results greatly indicate that simulated DM halos evolve to a maximum entropy state that is DARKexp. 

The results of this work are exciting as they show, for the first time, that the dynamical evolution of self-gravitating Newtonian systems follows the principle of the 2nd law of thermodynamics, where the entropy functional, given by $S_D$, increases over time. Future work should test DARKexp on larger sample sizes of equilibrium halos as well as on halos gathered from different simulations with varying halo evolution techniques.

\vspace{0.5cm}\noindent
\acknowledgments
\indent \indent We acknowledge the use of simulation data from the IllustrisTNG project. The IllustrisTNG simulations were undertaken with compute time awarded by the Gauss Centre for Supercomputing (GCS) under GCS Large-Scale Projects GCS-ILLU and GCS-DWAR on the GCS share of the supercomputer Hazel Hen at the High Performance Computing Center Stuttgart (HLRS), as well as on the machines of the Max Planck Computing and Data Facility (MPCDF) in Garching, Germany.
AF was supported by the College of Science and Engineering Fellowship at the University of Minnesota.
JH was supported by a research grant (VIL54489) from VILLUM FONDEN.

\bibliographystyle{unsrt}
\bibliography{main}

\end{document}